\documentclass{aa}  

\usepackage{natbib}
\usepackage{graphicx}
\usepackage{txfonts}
\usepackage{hyperref}
\hypersetup{colorlinks=true,citecolor=blue}
\usepackage{caption}
\usepackage{subcaption}
\newcommand{\Mp}{M_{\rm p}}
\newcommand{\Rp}{R_{\rm p}}
\newcommand{\Mpj}{M_{{\rm p}_{\rm j}}}
\newcommand{\Rpj}{R_{{\rm p}_{\rm j}}}

\newcommand{\pj}{{\rm p}_{\rm j}}

\newcommand{\taupj}{\tau_{{\rm p}_{\rm j}}}
\newcommand{\taup}{\tau_{{\rm p}}}
\newcommand{\taus}{\tau_{\star}}

\newcommand{\Opj}{\Omega_{\pj}}

\newcommand{\Cpj}{C_{{\rm p}_{\rm j}}}

\newcommand{\Msun}{M_{\odot}}
\newcommand{\Rsun}{R_{\odot}}
\newcommand{\Ms}{M_{\star}}
\newcommand{\Rs}{R_{\star}}

\newcommand{\Os}{\Omega_{\star}}
\newcommand{\kpj}{k_{2,{\rm p}_{\rm j}}}
\newcommand{\kfpj}{k_{2f,{\rm p}_{\rm j}}}

\newcommand{\kfp}{k_{2f,{\rm p}}}
\newcommand{\kp}{k_{2,{\rm p}}}

\newcommand{\ks}{k_{2,\star}}
\newcommand{\kfs}{k_{2f,\star}}
\newcommand{\rj}{r_{\rm j}}

\newcommand{\G}{\mathcal{G}}

\usepackage[normalem]{ulem}
\usepackage{color}
\definecolor{blue}{RGB}{0,0,255}
\definecolor{red}{RGB}{255,0,0}
\definecolor{green}{RGB}{0,200,0}
\definecolor{black}{RGB}{0,0,0}

\begin{document} 
  \title{On the impact of tides on the transit-timing fits to the TRAPPIST-1 system}
  \author{E. Bolmont\inst{1}
    \and
        B.-O. Demory\inst{2}
    \and
        S. Blanco-Cuaresma\inst{3}
    \and
        E. Agol\inst{4}
    \and
        S. L. Grimm\inst{2}
    \and
        P. Auclair-Desrotour\inst{2}
    \and
        F. Selsis\inst{5}
    \and
        A. Leleu\inst{2}}

  \institute{Observatoire de Gen\`eve, Universit\'e de Gen\`eve, 51 Chemin des Maillettes, CH-1290 Sauverny, Switzerland\\
              \email{emeline.bolmont@unige.ch}
         \and
             Physikalisches Institut, Universit\"at Bern, Gesellschaftsstrasse 6, CH-3012 Bern, Switzerland
        \and
            Harvard-Smithsonian Center for Astrophysics, 60 Garden Street, Cambridge, MA 02138, USA
        \and
            Astronomy Department, University of Washington, Seattle, WA 98195, USA
        \and
            Laboratoire d'astrophysique de Bordeaux, Universit\'e de Bordeaux, CNRS, B18N, all\'ee Geoffroy Saint-Hilaire, 33615, Pessac, France
             }

  \date{}

  \abstract{Transit Timing Variations, or TTVs, can be a very efficient way of constraining masses and eccentricities of multi-planet systems.
  Recent measurements of the TTVs of TRAPPIST-1 led to an estimate of the masses of the planets, enabling an estimate of their densities and their water content \citep{2018A&A...613A..68G}.
  A recent TTV analysis using data obtained in the past two years yields a 34\% and 13\% increase in mass for TRAPPIST-1b and c, respectively.
  In most studies to date, a Newtonian N-body model is used to fit the masses of the planets, while sometimes general relativity is accounted for. 
  Using the Posidonius N-body code, in this paper we show that in the case of the TRAPPIST-1 system, non-Newtonian effects might be also relevant to correctly model the dynamics of the system and the resulting TTVs. 
  In particular, using standard values of the tidal Love number $k_2$ (accounting for the tidal deformation) and the fluid Love number $k_{2f}$ (accounting for the rotational flattening) leads to differences in the TTVs of TRAPPIST-1b and c similar to the differences caused by general relativity. 
  We also show that relaxing the values of tidal Love number $k_2$ and the fluid Love number $k_{2f}$ can lead to TTVs which differ by as much as a few 10~s on a $3-4$-year timescale, which is a potentially observable level. 
  The high values of the Love numbers needed to reach observable levels for the TTVs could be achieved for planets with a liquid ocean, which, if detected, might then be interpreted as a sign that TRAPPIST-1b and TRAPPIST-1c could have a liquid magma ocean.
  For TRAPPIST-1 and similar systems, the models to fit the TTVs should potentially account for general relativity, for the tidal deformation of the planets, for the rotational deformation of the planets and, to a lesser extent, for the rotational deformation of the star, which would add up to 7x2+1 = 15 additional free parameters in the case of TRAPPIST-1.}


\maketitle


\section{Introduction}

The measurement of Transit Timing Variations (TTVs) in the context of multi-transiting planet systems can be a very efficient method to derive dynamical parameters of a planetary system, such as mass and eccentricity \citep[see][for a review]{2018haex.bookE...7A}.
The TRAPPIST-1 system has been intensely monitored by TRAPPIST, K2 and Spitzer, which led to estimates of the masses of the planets by \citet{2018A&A...613A..68G}.
Recently additional Spitzer observations were obtained thanks to the Spitzer proposal \#14223 \citep{2019sptz.prop14223A}.

In most studies on TTVs, the model used is an N-body model assuming point-mass/Newtonian dynamics, with sometimes a prescription for general relativity \citep[as in][]{2018A&A...613A..68G,2008ApJ...685..543J,2008MNRAS.389..191P}.  Theoretical studies have considered the possible impact of tides and quadrupole distortion on transit times \citep{2002ApJ...564.1019M,2007MNRAS.377.1511H}.
However, the influence of tides has never been consistently taken into account in a multi-planet context.

Some studies do take into account tidal decay \citep[e.g.][]{2018AcA....68..371M}, but decay typically occurs on timescales much longer than the typical duration of observations that are available for TRAPPIST-1.
However, tidal forces are not only a dissipative effect (which drives migration and spin evolution), there is also a non-dissipative part which depends on the real part of the complex Love number of degree 2, $k_2$, which quantifies the shape of the tidal deformation.
This deformation can lead to a precession of the orbit which can lead to TTVs .
In addition, for fast rotating planets, rotational flattening can also drive a precession of the orbit which can lead to TTVs. 
These effects have been considered in systems with a single hot-Jupiter planet \citep[see][for a comparative study of each effect]{2009ApJ...698.1778R}. 
The precession of the orbit leads to observable TTVs, which then can inform the internal structure of the planet through the determination of the Love number.

However, these effects are usually never taken into account when investigating the TTVs of multi-planetary systems. 
We show that in the context of TRAPPIST-1 \citep{2017Natur.542..456G,2016Natur.533..221G}, the inclusion of tidal forces may lead to an observable TTV signal. 
In contrast with TRAPPIST-1 d-h, planets b and c are in proximity to a higher-order resonance (increasing the frequency of the TTV pattern modulation; see \citealt{2005MNRAS.359..567A}) and exhibit small TTV amplitudes (2 to 5 min);  both effects inflate the uncertainties on the masses and eccentricities, as shown in \citet{2018A&A...613A..68G}.
Interestingly, a recent TTV analysis using data obtained in the past two years yields a 34\% increase in mass for TRAPPIST-1b and a 13\% increase in mass for TRAPPIST-1c (in preparation) compared to \citet{2018A&A...613A..68G}. These mass increases of the two inner planets drew our attention to physical processes that could impact the planet physical and orbital parameters on secular timescales. As the parameters for the other planets have remained relatively insensitive to the addition of new data, two hypotheses remain.
A first possibility is that the changing masses are due to an incomplete sampling of the TTV pattern that should resolve as new data are included. 
A second possibility is that dynamical models are missing physical processes which impact  the close-in planets more strongly, such as tides and rotational flattening. 

We show in this letter that the precession caused by general relativity, by tidal deformation and by rotational flattening could lead to significantly different TTVs for the two inner planets of TRAPPIST-1.


\section{Simulation set-up}\label{setup}

We use {\sc Posidonius}\footnote{\url{https://www.blancocuaresma.com/s/posidonius}} v2019.07.30 (\citealt{2017ewas.confE...8B}; Blanco-Cuaresma \& Bolmont, in prep), an N-body code which allows users to take into account additional forces and torques: tidal forces and torques, rotational flattening forces and torques, and general relativity \citep{2015A&A...583A.116B}.
{As in \citet{2015A&A...583A.116B}, tides are computed between a planet and the star independently of the other planets and the planet-planet tides are not taken into account \citep[which is justified, see][]{2019ApJ...875...22H}.} 
In Posidonius, we use the integrator IAS15 \citep{2015MNRAS.446.1424R} to compute the evolution of the system for 1500~days, which is approximately the time range available from all the observations collected from the system, and we fix the maximum timestep allowed to be 0.01~day~$= 14$~min. 
We tested the convergence of our code with timesteps of 0.005~day and 0.001~day, and find that the transit timings are stable to a precision of better than $10^{-6}$~s.

\subsection{Tidal model}

Posidonius enables accounting for equilibrium tides following the prescription of \citet{2015A&A...583A.116B}, which is an implementation of the constant-time-lag model \citep{1979M&P....20..301M, 1981A&A....99..126H, 1998ApJ...499..853E}.
{The equilibrium tide is the result of the hydrostatic adjustment of a body, in contrast to the dynamical tide which is the tidal response corresponding to the propagation of waves (e.g. inertial waves in the convective region of stars, see \citealt{1975A&A....41..329Z}; or gravito-inertial waves in a planetary liquid layer, see \citealt{2019A&A...629A.132A}).}

We review here the expressions for the tidal force and torques.
Let us consider a star, defined by its mass $\Ms$, its radius $\Rs$, its degree 2 potential Love number $\ks$, its (constant) time lag $\Delta\taus$, and its spin vector $\mathbf{\Os}$. 
Let us consider one planet, j, orbiting the star at a distance $\rj$. 
The planet is defined by its mass $\Mpj$, its radius $\Rpj$, its degree 2 potential Love number $\kpj$, its (constant) time lag $\Delta\taupj$, and its spin vector $\mathbf{\Opj}$.

Let us define $\mathbf{F_{{\rm diss},\pj}}$ and $\mathbf{F_{{\rm nodiss},\pj}}$ as the dissipative part and the non-dissipative part, respectively, of the force exerted on planet j due to the planetary tide as \citep{2015A&A...583A.116B} 
\begin{align}
\mathbf{F_{{\rm nodiss},\pj}}   & = \frac{-3\G}{\rj^7}\Ms^2\kpj\Rpj^5 \mathbf{e_{\rj}},\label{F_things_pl_1} \\
\mathbf{F_{{\rm diss},\pj}}     & = -9\G\frac{\dot{\rj}}{\rj^8}\Ms^2\Rpj^{5}\kpj \Delta\taupj \mathbf{e_{\rj}} \notag\\
    & \quad + 3\G\frac{\Ms^2\Rpj^{5}}{\rj^7}\kpj\Delta\taupj \left(\mathbf{\Opj}- \dot{\boldsymbol{\theta}_{\rm j}}\right)\times \mathbf{e_{\rj}} \label{F_things_pl_2},
\end{align}
where $\mathbf{e_{\rj}}$ is the unit vector $\mathbf{\rj}/\rj$ and $\dot{\boldsymbol{\theta}_{\rm j}}$ is a vector collinear with the orbital angular momentum of planet j, the norm of which is equal to the time derivative of the true anomaly. 
Let us define $\mathbf{F_{{\rm diss},\star}}$ and $\mathbf{F_{{\rm nodiss},\star}}$ as the dissipative part and the non-dissipative part, respectively, of the force exerted on planet j due to the stellar tide as
\begin{align}
\mathbf{F_{{\rm nodiss},\star}} & = \frac{-3\G}{\rj^7}\Mpj^2\ks\Rs^5 \mathbf{e_{\rj}}, \label{F_things_st_1}\\
\mathbf{F_{{\rm diss},\star}}   & = -9\G\frac{\dot{\rj}}{\rj^8}\Mpj^2\Rs^{5}\ks \Delta\taus \mathbf{e_{\rj}}\notag \\
    & \quad + 3\G\frac{\Mpj^2\Rs^{5}}{\rj^7}\ks\Delta\taus \left(\mathbf{\Os} - \dot{\boldsymbol{\theta}_{\rm j}} \right) \times \mathbf{e_{\rj}} \label{F_things_st_2}.
\end{align}
The total force as a result of the tides acting on a planet j is therefore given by the sum of these contributions \citep{2015A&A...583A.116B}
\begin{equation}\label{acc}
\begin{split}
\mathbf{F^{T}_{\pj}} &= \mathbf{F_{{\rm diss},\pj}} + \mathbf{F_{{\rm nodiss},\pj}} + \mathbf{F_{{\rm diss},\star}} + \mathbf{F_{{\rm nodiss},\star}}.
\end{split}
\end{equation}

\subsection{Rotational flattening model}

To account for rotational flattening, we also follow here the prescription of \citet{2015A&A...583A.116B}, which assumes that the deformation due to the rotational flattening results in a triaxial ellipsoid symmetric with respect to the rotation axis \citep{1999ssd..book.....M}.
This deformation is quantified by a parameter, $J_2$, which depends on the radius, mass and spin of the body and on the potential Love number of degree 2 for a perfectly fluid body (which we call here the fluid Love number, \citealt{2013ApJ...767..128C}).
We define this parameter for a planet j and the star as follows
\begin{align}
J_{2,{\rm p_j}} &=  \kfpj \frac{\Opj^2\Rpj^3}{3\G\Mpj}, \label{J2p} \\
J_{2,\star} &=  \kfs \frac{\Os^2\Rs^3}{3\G\Ms}. \label{J2s}
\end{align}
Let us define $\mathbf{F_{{\rm rot},\pj}}$ the force exerted on planet j due to the rotational flattening of planet j and $\mathbf{F_{{\rm rot},\star}}$ the force exerted on planet j due to the rotational flattening of the star as \citep{1999ssd..book.....M,2011CeMDA.111..105C}
\begin{align}
\mathbf{F_{{\rm rot},\pj}} &= \left(-\frac{3}{\rj^5}\Cpj+ \frac{15}{\rj^7}\Cpj \frac{\left(\mathbf{\rj}.\mathbf{\Opj}\right)^2}{\Opj^2}\right)\mathbf{\rj} 
- \frac{6}{\rj^5}\Cpj \frac{\mathbf{\rj}.\mathbf{\Opj}}{\Opj^2}\mathbf{\Opj}, \label{acc_rot_pl_1}\\
\mathbf{F_{{\rm rot},\star}} &= \left(-\frac{3}{\rj^5}C_\star+ \frac{15}{\rj^7}C_\star \frac{\left(\mathbf{\rj}.\mathbf{\Os}\right)^2}{\Os^2}\right)\mathbf{\rj} - \frac{6}{\rj^5}C_\star \frac{\mathbf{\rj}.\mathbf{\Os}}{\Os^2}\mathbf{\Os}, \label{acc_rot_pl_2}
\end{align}
where $C_\star$ and $\Cpj$ are defined as follows: 
\begin{align}
  \Cpj &= \frac{1}{2}\G\Mpj\Ms J_{2,{\rm p_j}}\Rpj^2, \label{C_rot_1}\\
C_\star &= \frac{1}{2}\G\Mpj\Ms J_{2,\star}\Rs^2. \label{C_rot_2}  
\end{align}

The resulting force on planet j due to the rotational deformation of both the star and planet j is the sum of both contributions 
\begin{equation}\label{acc_rot}
\mathbf{F^{R}_{\pj}} = \mathbf{F_{{\rm rot},\pj}} + \mathbf{F_{{\rm rot},\star}}.
\end{equation}

\subsection{General Relativity}\label{GenRel}

We use three different prescription for general relativity: \citet{1995PhRvD..52..821K}, which is the one used in Mercury-T \citep{2015A&A...583A.116B}; \citet{1975ApJ...200..221A}; and \citet{1983A&A...125..150N}.

The prescription of \citet{1995PhRvD..52..821K} was designed for 2 bodies and Posidonius takes into account the post-Newtonian, the spin-orbit, and the post$^2$-Newtonian contributions to the total acceleration (Eqs. 2.2b, 2.2c, 2.2d, respectively), as well as the spin precession equations for both bodies (Eqs.~2.4a and 2.4b).

The prescription of \citet{1975ApJ...200..221A} accounts for the post-Newtonian acceleration of two bodies.
We refer the reader to Eq.~12 of \citet{1975ApJ...200..221A} where the expression of this acceleration is given.

The prescription of \citet{1983A&A...125..150N} is more complete in so far as it accounts for the post-Newtonian effect between all bodies. 
We refer the reader to Eq.~1 of \citet{1983A&A...125..150N} which gives the point-mass acceleration.
Posidonius accounts for this acceleration, except for the last term which accounts for the perturbation of 5 solar system asteroids. 

\section{Transit Timing Variations}

We perform simulations of the TRAPPIST-1 system switching on and off these various effects: the effect of the planetary tide (by varying the Love number $\kp$ and the time lag $\taup$), the effect of the stellar tide (by varying the Love number $\ks$), the effect of the rotational flattening of the planets (by varying the fluid Love number $\kfp$), the effect of the rotational flattening of the star (by varying the fluid Love number $\kfs$) and the effect of general relativity. 
We list in Table~\ref{table:1} the reference values of the parameters which we vary here and we refer the reader to Appendix~\ref{A1}, which lists the parameters which remain constant in our simulations, which include the initial orbital elements for the planets.
We tested the three different prescriptions of the general relativity introduced in Section~\ref{GenRel}.
They gave very similar results so that in the following we compare the other effects with respect to the simulations performed using the prescription of \citet{1995PhRvD..52..821K}.

The planetary reference values were taken to be representative of the Earth; in particular the quantity $\kp\Delta\taup$ is equal to 213~s \citep{1997A&A...318..975N}.
The stellar reference values were chosen to be representative of fully convective M-dwarfs \citep{2015A&A...583A.116B}.

\begin{table}[htbp]
\centering                          
\begin{tabular}{c c c}        
\hline\hline                 
Parameter       & value      \\    
\hline                        
$\kfp^{{\rm ref}}$              & 0.9532     \\ 
$\kp^{{\rm ref}}$               & 0.299      \\
$\Delta\taup^{{\rm ref}}$ (s)   & 712.37     \\ 
\hline
$\kfs^{{\rm ref}}$              & 0.307      \\
$\ks^{{\rm ref}}$               & 0.307      \\
\hline
\end{tabular}
\caption{Reference values for the parameters we vary in this study. We assume that all planets of the system have the same potential Love number of degree 2 $\kfp$, fluid Love number $\kp$ and time lag $\Delta\taup$.}\label{table:1}      
\end{table}

For all the simulations we performed, we calculated the transit timing variations \citep{2018haex.bookE...7A} as follows: (i) for each transit, we find the time of the transit mid-time by performing an interpolation to find the precise time a given planet crosses a reference direction. This corresponds to the ``Observed transit time'' \textit{O}; (ii) we evaluate the ``Calculated transit times'' \textit{C} by performing a linear fit\footnote{We use the function \texttt{LinearRegression} from the \texttt{linear\_model} package of scikit-learn.} of the transit times calculated in step 1) over the total number of transit; (iii) we calculate the difference \textit{O}-\textit{C} to obtain the TTVs as a function of the epoch (or transit number).

To quantify the impact of each additional effect on the simulated TTVs, we compute the difference between the TTVs calculated taken into account an additional effect and the TTVs obtained for a Newtonian N-body integration.

\section{Influence of each effect on the TTVs}\label{all_effects}

   \begin{figure*}
   \centering
   \includegraphics[width=1\linewidth]{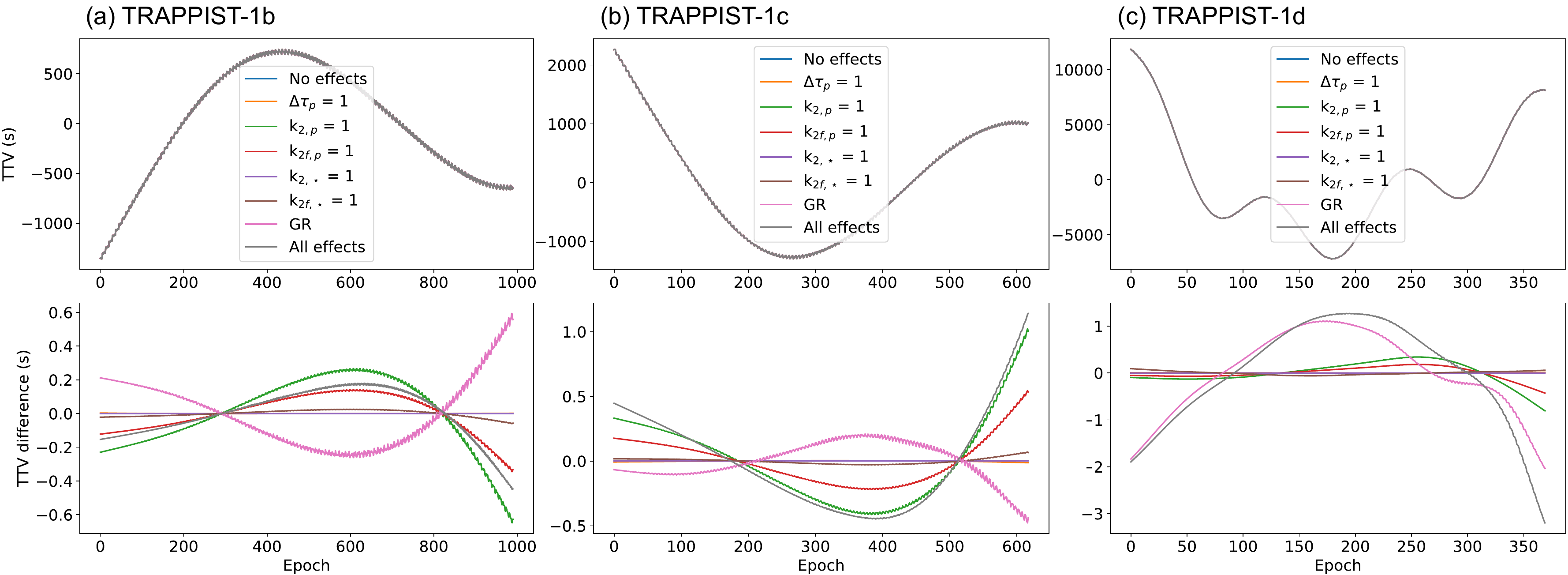}
   \caption{Impact of various additional effects on the TTVs of (a) planet b, (b) planet c and (c) planet d. Top panel: Transit Timing Variations for a pure N-body simulation (blue), for a simulation for which $\Delta\taup = 1\times\Delta\taup^{{\rm ref}}$ for all planets (orange), for a simulation for which $\kp = 1 \times \kp^{{\rm ref}}$ for all planets (green); for a simulation for which $\kfp = 1 \times \kfp^{{\rm ref}}$ for all planets (red); for a simulation for which $\ks = 1 \times \ks^{{\rm ref}}$ (purple); for a simulation for which $\kfs = 1 \times \kfs^{{\rm ref}}$ (brown); for a simulation where we only consider the general relativity (pink) and for a simulation for which all effects are taken into account. Bottom panel: the corresponding TTVs differences with the Newtonian N-body case.}
    \label{fig:1}%
    \end{figure*}

We performed a set of 6 simulations of the TRAPPIST-1 system to test the impact of the additional effects listed in Section~\ref{setup}, and we compare each with a Newtonian N-body simulation. 
One after another, we explored the effect of each parameter using the reference values of Table~\ref{table:1} and general relativity. 
We tested the influence of the dissipative part of the planetary tide by assuming $\Delta\taup = 1\times\Delta\taup^{{\rm ref}}$ for all planets, with all other parameters set to zero.
We repeated the operation for the non-dissipative part of the planetary tide (through $\kp$, equal for all planets), the rotational flattening of the planets (through $\kfp$, equal for all planets), the non-dissipative part of the stellar tide (through $\ks$), the rotational flattening of the star (through $\kfs$), and for general relativity.

Figure~\ref{fig:1} shows the results for planets b to d. 
The top panels of Fig.~\ref{fig:1} show the transit timing variations for the 3 planets for the 7 simulations and the bottom panels show the difference between the TTVs and the TTVs corresponding to the pure N-body simulation.
The different additional effects have a very limited impact on the shape of the TTVs, but computing the difference with the pure N-body case reveals the amplitude of each effect.


For TRAPPIST-1b (T-1b), the dominant effects are the non-dissipative part of the planetary tide (green in Figure~\ref{fig:1}a), and general relativity (pink), respectively accounting for a difference in TTVs of about -0.63~s and 0.56~s at the end of the 1500 day. 
The effect of the rotational flattening of the planets (red) plays a smaller role but still accounts for more than half the amplitude due to the non-dissipative part of the planetary tide with a difference of -0.33~s.
The effect of the dissipative part of the planetary tide (orange) and the non-dissipative part of the stellar tide (purple) are completely negligible (accounting for a difference $\sim 1\times 10^{-3}$~s and $\sim 1\times 10^{-4}$~s respectively), which is in agreement with \citet{2009ApJ...698.1778R}.
The effect of the rotational flattening of the star (brown) is much smaller (accounting for a difference of -0.057~s), but might contribute at a lesser extent.
Accounting for all effects (grey) leads to an absolute difference of -0.45~s at the end of the 1500 day simulation. 
The effects of general relativity and the non-dissipative part of the planetary tide almost cancel each other out, while the amplitude is determined by the effect of the rotational flattening of the planet (red curve) and of the star (brown curve).


For TRAPPIST-1c (T-1c), the dominant effects are the non-dissipative part of the planetary tide (accounting for a difference of 1~s, in green in Fig.\ref{fig:1}b), followed by the effect of the rotational flattening of the planet (0.53~s, in red), followed by the effect of general relativity (-0.44~s, in pink). 
The effect of the rotational flattening of the star accounts for 0.067~s (purple) and the effect of the dissipative part of the planetary tide remains negligible (0.013~s, orange).
As with T1-b, to reproduce the difference observed when all effects are taken into account (in grey in Fig.\ref{fig:1}b), one needs to account for the non-dissipative part of the planetary tide, the rotational flattening of the planet, general relativity and the rotational flattening of the star to a lesser extent.
Note that the precession of the orbits due to the rotational flattening depends on the square of the spin frequency of the considered body \citep{2009ApJ...698.1778R}. 
Here we use a rotation period of 3.3~days for TRAPPIST-1 \citep{2017NatAs...1E.129L}. It is possible that the rotation is slower \citep[as the period distribution of nearby late M dwarfs shows,][]{2016ApJ...821...93N}, in which case, the contribution of the rotational flattening of the star would be even less important.


For TRAPPIST-1d (T-1d), Fig.~\ref{fig:1}c) shows that the dominant effect is general relativity (-2.03~s, in pink).
The effect of the non-dissipative part of the planetary tide (accounting for -0.81~s, in green) and the effect of the rotational flattening of the planet (accounting for -0.43~s, in red) should also probably be taken into account.
This is also true for all the external planets: the amplitude due to general relativity is much higher but at the same time not accounting for at least the non-dissipative part of the planetary tide and the rotational flattening of the planet leads to small offsets (see Appendix~\ref{A2}, Fig~\ref{figA2:1}). 

\section{Potential observable effects}

   \begin{figure*}
   \centering
   \includegraphics[width=0.8\linewidth]{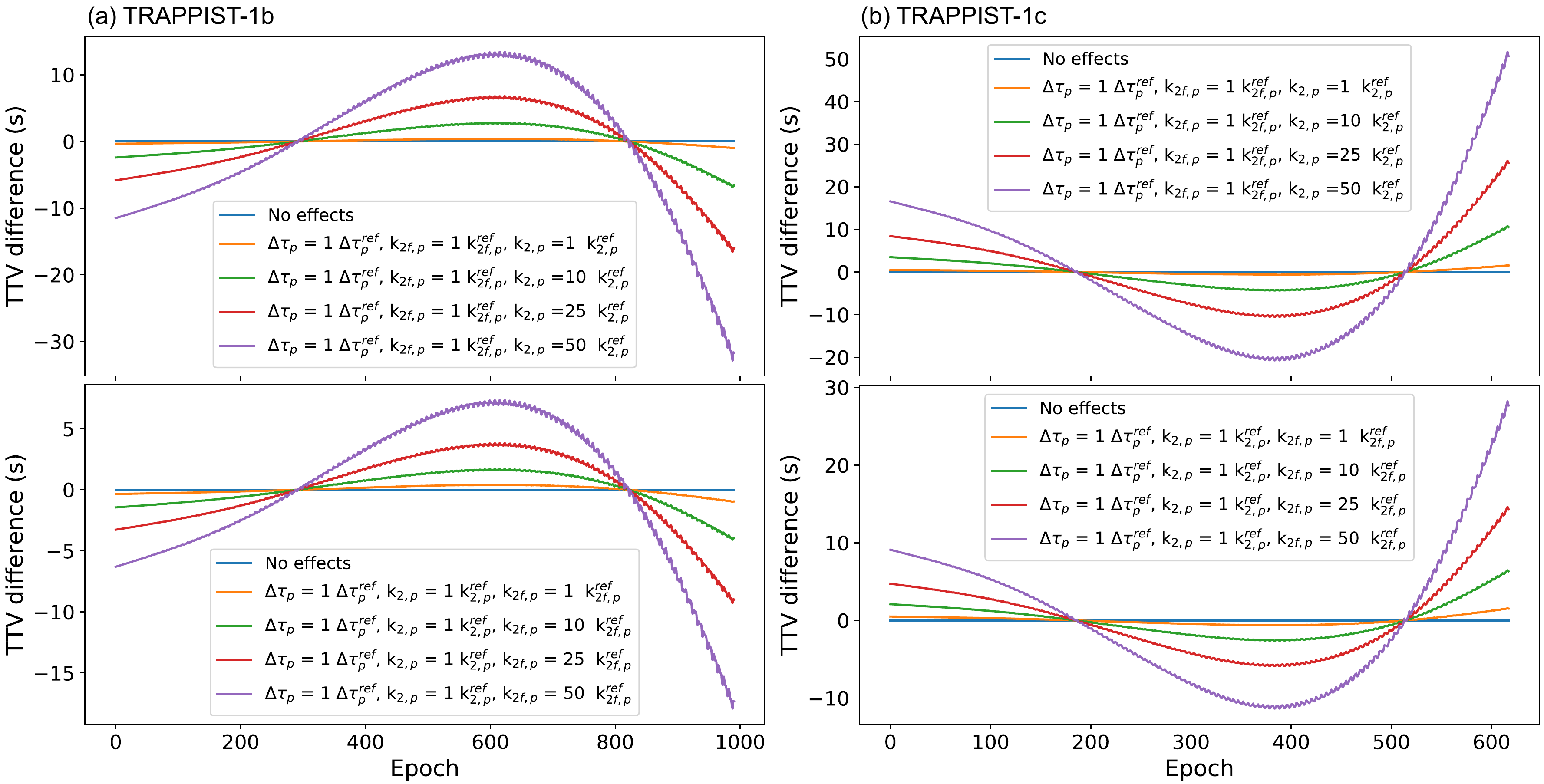}
   \caption{TTVs differences with the pure N-body case for (a) planet b and (b) planet c. Top panel: fixing the planetary fluid Love number and the dissipation to their reference values (see Table~\ref{table:1}), the potential Love number $\kp$ is varied between 1 and 50$\times \kp^{{\rm ref}}\sim 15$. Bottom panel: fixing the planetary Love number and the dissipation to their reference values, the potential Love number $\kfp$ is varied between 1 and 50$\times \kfp^{{\rm ref}} \sim 48$.}
    \label{fig:2}%
    \end{figure*}

We performed simulations for which we varied the potential Love number $\kp$ and the fluid Love number $\kfp$ over a wide range.
We first treat these parameters as free parameters with no limitations on their value and then we will discuss in Section~\ref{discussion} the validity of this approach. 

As in the previous Section~\ref{all_effects}, we always assume the same value of the Love numbers for all planets. 
We vary the parameter $\kp$ from $1\times\kp^{{\rm ref}} = 0.299$ to $50\times\kp^{{\rm ref}} = 14.95$, and the impact on the TTV differences with the pure N-body case can be seen in the top panel of Figure~\ref{fig:2}a) for TRAPPIST-1b and in the top panel of Figure~\ref{fig:2}b) for TRAPPIST-1c.


Considering the highest value of the tidal Love number leads to a difference in TTVs after 1500 day of -31.7~s for T-1b and 50.7~s for T-1c. 
The amplitude of these effects are then comparable to the precision achievable today on the observed TTVs of the two inner planets. 
If the tidal Love number could reach such high values, the effect of the non-dissipative part of the planetary tide (the tidal deformation of T-1b and T-1c) could therefore be detectable.

Similarly, we vary the parameter $\kfp$ from $1\times\kfp^{{\rm ref}} = 0.9532$ to $50\times\kfp^{{\rm ref}} = 47.66$, and the impact on the TTV differences with the pure N-body case can be seen in the bottom panel of Figure~\ref{fig:2}a) for TRAPPIST-1b and in the bottom panel of Figure~\ref{fig:2}b) for TRAPPIST-1c.
Assuming the highest value of the fluid Love number leads to a difference in TTVs after 1500 day of -17.3~s for T-1b and 27.7~s for T-1c. 

Considering all important effects given of Section~\ref{all_effects} and relaxing the range of possible Love numbers might therefore be a way to settle the question of the increasing masses of T-1b and T-1c and to settle the two hypotheses given in the introduction: is it a sampling problem or are we missing dynamical processes?  
The answer to this question depends upon the potential degeneracy of these effects with varying the N-body parameters, as well as the duration and precision of the transit timing measurements.  
We do not explore these effects in this letter.

\section{Discussion}\label{discussion}



We showed that for systems like TRAPPIST-1, the effect of the tidal deformation of the planets (through the planetary tidal Love number $\kp$), the effect of the rotational deformation of the planets (through the planetary fluid Love number $\kfp$) and the effect of the rotational flattening of the star (through the stellar fluid Love number $\kfs$) can impact the TTVs of the 2 inner planets at the same order of magnitude as the general relativity if we assume ``standard'' values for these parameters. 
We also showed that the tidal dissipation (responsible for the misalignment which drives long term tidal evolution) is not significantly impacting the TTVs of the system over the short observation time that we simulated.
By relaxing the assumptions on the planetary tidal and fluid Love numbers, we also showed that a high Love number can lead to differences in TTV of the order of $\sim 10$~s. 
This difference is potentially observable with the current precision we have on the transit timings, unless there is significant degeneracy with other N-body parameters.

However, it is commonly accepted that a tidal Love number cannot exceed 1.5, which corresponds to a homogeneous body. 
Which means that the physical range of our study should encompass at maximum values which are $5\times\kp^{{\rm ref}} = 5 \times 0.299$.
Limiting ourselves to this value would entail a difference in TTVs for T-1b of less than 2.5~s, which is below the precision we can achieve today.
On the other hand, it is known that if a planet has a liquid layer (liquid water ocean, or liquid magma ocean), the response of the body becomes more complex: in particular it becomes highly dependent on the excitation frequency.
Specifically, if a frequency excites a resonant mode of the ocean, the tidal response can be much higher than what a homogeneous-rocky-planet model would predict \citep[see for instance][]{2019A&A...629A.132A}.
{Investigating this aspect consistently will require to generalize the tidal formalism used in this letter to account for the frequency dependence of the dynamical tide \citep[e.g. use the formalism of][]{1961GeoJ....5..104K}.}

That is why, we think we might need to perform a TTV analysis of the TRAPPIST-1 system accounting for the various physical processes described in this letter, with no particular preconception about the values of the parameters for the planetary Love numbers. 
If the TTVs are reproduced by having a TRAPPIST-1b planet with a high Love number, this could be a sign for a liquid layer on the planet, possibly a magma ocean given the flux it receives and the tidal heat flux it might generate \citep[e.g.][]{2018A&A...612A..86T,2018ApJ...857..142M}.

{While a difference of a few $\sim 10$~s is potentially observable, it  could be interpreted as a system with slightly different planetary masses and periods by a classical TTV retrieval code.
}
{Our group is thus currently investigating if these effects could be picked up with such a retrieval code, and, if so, in which conditions (duration of the observations, precision of the timings). We are also working on implementing these effects in the TTV analysis pipeline and plan to revisit the analysis with the additional parameters mentioned earlier (Grimm et al., in prep).}

\begin{acknowledgements}
The authors would like to thank the anonymous referee for helping improving the manuscript.
This work has been carried out within the framework of the NCCR PlanetS supported by the Swiss National Science Foundation.
B.-O.D. acknowledges support from the Swiss National Science Foundation (PP00P2\_163967).
P. Auclair-Desrotour acknowledges financial support from the European Research Council via the Consolidator grant EXOKLEIN (grant number 771620).
F. S. acknowledges support from the CNRS/INSU PNP (Programme National de Planétologie).
This research has made use of NASA's Astrophysics Data System.
\end{acknowledgements}

\bibliographystyle{aa} 
\bibliography{Biblio} 

\begin{appendix} 
\section{Initial conditions for the TRAPPIST-1 simulations}\label{A1}
To ensure the reproducibility of our simulations, we give here the exact initial conditions we took for the system.
Table~\ref{tableA:1} gives the stellar parameters used for the integration of the system.
The stellar mass and radius come from \citet{2017Natur.542..456G} and the rotation comes from \citet{2017NatAs...1E.129L}.
The value of the radius of gyration squared $rg^2_\star$ \citep{1981A&A....99..126H} comes from \citet{2015A&A...583A.116B} and should be typical of a fully convective dwarf.

Table~\ref{tableA:2} gives the masses and radii of the planets as well as the initial orbital elements.
We consider that all planets have the same radius of gyration squared $rg^2_{\rm p} = 0.3308$ (where this quantity is related to the moment of inertia $I_{\rm p} = \Mp (\Rp rg_{\rm p})^2$).
We consider that all planets have a zero obliquity (angle between the direction perpendicular to the orbital plane and the rotation axis of the planet) and are tidally locked \citep[see discussion in][]{2017NatAs...1E.129L}.

To perform the integration of the system, we used {\sc Posidonius} v2019.07.30 (\url{https://www.blancocuaresma.com/s/posidonius}).
This version was slightly altered to be able to fix a maximum timestep size (0.01~day).
The initial conditions can be found in \citet{bolmont_emeline_2020_3634640}\footnote{\url{https://zenodo.org/record/3634640}}.

\begin{table}[htbp]
\caption{Stellar parameters}             
\label{tableA:1}      
\centering                          
\begin{tabular}{c c c}        
\hline\hline                 
Parameter       & value & unit \\    
\hline                        
$\Ms$           & 0.08  & $\Msun$      \\      
$\Rs$           & 0.117 & $\Rsun$\\
$rg^2_\star$     & 0.2   &\\ 
Rotation period & 3.3   & day\\
\hline                                   
\end{tabular}
\end{table}
\begin{table*}[h]
\caption{Masses, radii and initial orbital elements used for the dynamical simulations of the TRAPPIST-1 system. The inclination is here given with respect to the equatorial plane of the star.}             
\label{tableA:2}      
\centering          
\begin{tabular}{c c c c c c c c c }     
\hline\hline       
Planet  & Mass  & Radius    & Semi-major    & Eccentricity  & Inclination   & Mean      & Argument of   & Longitude of \\
        &       &           & axis          &               &               & anomaly   & pericenter    & ascending node \\
    & ($\Msun$) & ($\Rsun$) & (au)  &   & (degree)    & (degree)    & (degree)    & (degree) \\
\hline                    
   b & 2.97733e-06  & 1.127 & 0.01110318    & 0 & 0.59  & 90.0000000    & 0     & 0\\  
   c & 3.34950e-06  & 1.100 & 0.01520668    & 0 & 0.50  & 51.5815880    & 0     & 0\\
   d & 9.16102e-07  & 0.788 & 0.02142513    & 0 & 0.30  & 84.5759410    & 0     & 0\\
   e & 2.34751e-06  & 0.915 & 0.02815839    & 0 & 0.40  & 305.455247    & 0     & 0 \\
   f & 2.69105e-06  & 1.052 & 0.03705241    & 0 & 0.08  & 283.559942    & 0     & 0\\
   g & 3.29224e-06  & 1.154 & 0.04508048    & 0 & 0.37  & 233.773520    & 0     & 0\\
   h & 9.73359e-07  & 0.777 & 0.05955922    & 0 & 0.20  & 1.31390800    & 0     & 0\\
\hline                  
\end{tabular}
\end{table*}

\section{Transit Timing Variations for the 4 outer planets of TRAPPIST-1}\label{A2} 

\begin{figure}[htbp]
   \centering
   \includegraphics[width=0.92\linewidth]{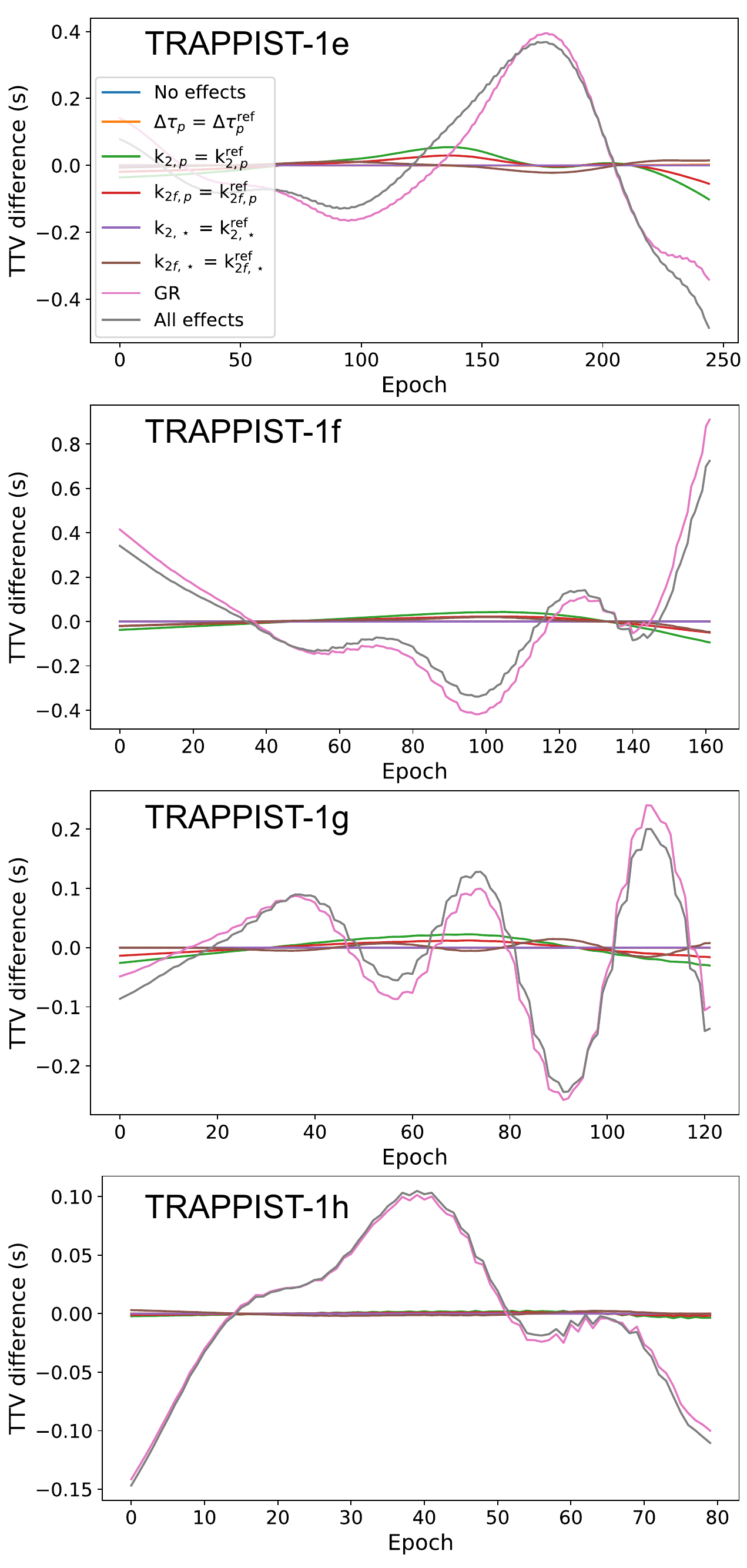}
   \caption{As the bottom panels of Fig~\ref{fig:1} but for a) TRAPPIST-1e, b) TRAPPIST-1f, c) TRAPPIST-1g and d) TRAPPIST-1h. The general relativity is the dominant effect but the planetary deformation (due to tides or rotation) is not quite completely negligible, except for TRAPPIST-1h.}
    \label{figA2:1}%
    \end{figure}

As in Figure~\ref{fig:1}, Figure~\ref{figA2:1} shows the difference in TTVs with the N-body case for 7 different simulations (N-body and simulations with additional effects) for TRAPPIST-1e to TRAPPIST-1h.
General relativity is the dominant effect but the non-dissipative part of the planetary tidal force (via $\kp$) and the rotational flattening of the planets (via $\kfp$) are still contributing marginally.
Only for TRAPPIST-1h is general relativity the only relevant process to account for. 

\end{appendix}

\end{document}